\documentstyle[aps,floats,epsf]{revtex}

\makeatletter 
\def\preprint#1{%
   \def\@preprint{\noindent\hfill\hbox{#1}\vskip 10pt}%
}
\makeatother
\draft

\preprint{
   \begin{tabular}{l}
     hep-ph/9907498 \\
     PSU/TH/210 \\
     DESY 99-092
   \end{tabular}
}

\title
      {Transversity Distribution Does Not Contribute to Hard Exclusive
       Electroproduction of Mesons} 

\author{J.C. Collins}
\address{Physics Department,
        Penn State University,
        104 Davey Laboratory,
        University Park PA 16802,
        U.S.A.
}

\author{M. Diehl}
\address{Deutsches Elektronen-Synchrotron DESY, 22603 Hamburg, Germany}

\date{26 July 1999}

\begin{document}

\twocolumn[\hsize\textwidth\columnwidth\hsize\csname@twocolumnfalse\endcsname
\maketitle

\begin{abstract}%
  We show that in hard exclusive electroproduction, $ep\to eVp$, the
  leading-twist hard-scattering coefficient for the production of a
  transversely polarized vector meson $V$ vanishes to all orders of
  perturbation theory.  This implies that this process cannot be used
  to measure the skewed transversity distribution of quarks in a
  hadron.  In contrast, a recent calculation obtained a non-zero value
  at NLO.  We show that this calculation is incorrect because it
  failed to include the necessary collinear subtractions.  Our method
  of proof also applies to other processes whose hard-scattering
  coefficients are constrained by chirality and helicity conservation,
  and thus validates helicity selection rules based on these
  symmetries.
\end{abstract}

\pacs{12.38.Bx, 12.38.Qk, 13.60.Fz, 13.60.Le}

\vskip 2pc]

\section{Introduction}
\label{intro}

There has been much interest recently in hard exclusive
electroproduction of vector mesons, $\gamma^* p\to V p$, at large
photon virtuality and small invariant momentum transfer to the proton.
In their paper \cite{cfs} proving factorization for this process,
Collins, Frankfurt and Strikman observed that the process appears to
provide a new probe of the (skewed) transversity distribution of
quarks in a proton, since this distribution appears in the
factorization formula for production of transversely polarized vector
mesons by longitudinally polarized virtual photons.  Unfortunately,
the lowest order coefficient function for this process vanishes, as
found by Mankiewicz, Piller and Weigl \cite{mpw}. Later Diehl, Gousset
and Pire \cite{dgp} showed how to extend this result to all orders in
$\alpha_{s}$ by the use of the chiral symmetry of massless QCD and of
rotational invariance.

But chiral invariance is broken by the axial anomaly of QCD, so
Hoodbhoy and Lu \cite{hl} proposed that this would permit a non-zero
coefficient function in non-leading order.  Moreover, they found that
in dimensional regularization the one-loop graphs for the process do
indeed violate chirality conservation, and so they appear to obtain a
non-zero coefficient function.  They derived their result in
$4-\epsilon$ dimensions from multiplying $1/\epsilon$ factors due to
divergences in the loop integrals by a factor of $\epsilon$ for the
violation of chiral symmetry in the lowest order coefficient function.
If their calculation is correct, it shows that we have a new process
for measuring transversity distributions.  This result would be
important because of the paucity of leading twist processes that are
sensitive to transversity distributions.

We show in this paper that their conclusion is, in fact, mistaken, and
that the coefficient function does indeed vanish.  The problem with
the calculation of Ref.\ \cite{hl} is that to compute a one-loop
contribution to a coefficient function from the corresponding graphs,
one must subtract the zero-loop coefficient function times one-loop
contributions to the parton distributions and distribution amplitudes.
The subtractions exactly cancel the chirality violating term and
result in a coefficient function that is zero.  We explain this in
detail in Sec.\ \ref{one.loop}.

Then, in Sec.\ \ref{proof} we generalize this result to all orders of
perturbation theory.  We complete the proof in Ref.\ \cite{dgp} that
the coefficient function vanishes.  The vanishing is a consequence of
properties of QCD that are related to chiral symmetry and remain true
in the dimensionally regulated theory.  In Sec.\ \ref{anomaly} we
comment on the relation between our result and the existence of the
axial anomaly.

There are several other situations in perturbative QCD where chiral
invariance is used to restrict the polarization states that occur at
leading-twist.  One example is hadronic helicity conservation
\cite{bl}, which holds in a number of exclusive processes. It applies
for instance to the electromagnetic form factors of $\rho$-mesons,
whose hard-scattering coefficient is closely related to the one of
exclusive meson production \cite{dgp}.  The objection of Hoodbhoy and
Lu applies to these situations; the proofs previously given in the
literature appear to be invalidated by the axial anomaly.  The methods
given in the present paper correct these proofs and show that the
results obtained by chiral symmetry applied to hard-scattering
coefficients are indeed valid.

Similarly, it is well known, on the basis of chirality conservation
arguments, that chiral odd parton densities and fragmentation
functions occur in pairs but not singly \cite{chiral.odd}.  Again the
methods presented in the present paper ensure that these results are
valid despite the anomalous violation of chiral symmetry.

\section{One-loop calculation with subtractions}
\label{one.loop}

\begin{figure}
   \begin{center}
        \leavevmode
        \epsfxsize=0.34\hsize
        \epsfbox{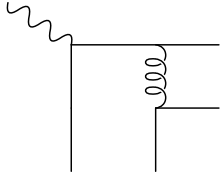}
   \end{center}
   \caption{A lowest order graph for the hard-scattering coefficient.}
   \label{fig.LO}
\end{figure}

If the calculation of the coefficient function for our process could
be performed in massless QCD without a regulator, then one could
immediately apply the proof of Ref.\ \cite{dgp} to show that the
coefficient function vanishes.  But there are collinear divergences,
so a regulator must be applied.  When dimensional regularization with
$4-\epsilon$ dimensions is used, it is easy \cite{hl} to see that tree
graphs, like Fig.\ \ref{fig.LO}, for the coefficient function for our
process are proportional to $\epsilon$.  When the graph is contracted
with chiral odd quark distribution amplitudes, one obtains a factor of
the form
\begin{equation}
   \gamma^{\lambda} \gamma^{+}\gamma^{i}\gamma_{\lambda},
\label{contraction}
\end{equation}
where $\lambda$ is summed over and $i$ is a transverse index.  Here
and in the following we use light-cone notation, $\gamma^{\pm} =
(\gamma^0 \pm \gamma^3) /\sqrt{2}$.  The matrix in (\ref{contraction})
equals $- \epsilon \gamma^{+} \gamma^{i}$, and hence in 4 dimensions
the coefficient function is zero.

\begin{figure}
   \begin{center}
        \leavevmode
        \epsfxsize=0.34\hsize
        \epsfbox{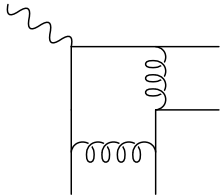}
   \end{center}
   \caption{An NLO graph for the hard-scattering coefficient.}
   \label{fig.NLO}
\end{figure}

Consider now a one-loop graph, such as Fig.\ \ref{fig.NLO}.  It has
collinear divergences when one or other of the gluons becomes parallel
to the attached quark lines.  The resulting $1/\epsilon$ multiplies a
factor $\epsilon$ of the same origin as in a tree graph and gives a
nonzero result at $\epsilon=0$.  Hoodbhoy and Lu \cite{hl} computed
the sum over all graphs and found that the result remains non-zero.

However, this does not represent the full calculation of the
coefficient function.  Schematically the factorization theorem is
\begin{equation}
    A = H \ast f \ast \phi ,
\label{factorize}
\end{equation}
where the star denotes convolution, $A$ is the amplitude for the
process, up to power corrections in the photon virtuality, $H$ is the
hard-scattering function, $f$ is the skewed parton distribution in the
target, and $\phi$ is the meson distribution amplitude.  Let us apply
the theorem to massless quark states and take the one-loop terms (of
order $\alpha_{s}^{2}$).  We have
\begin{equation}
   A_{1} = H_{1} \ast f_{0} \ast \phi_{0} 
           + H_{0} \ast f_{1} \ast \phi_{0} 
           + H_{0} \ast f_{0} \ast \phi_{1} ,
\end{equation}
where the subscripts represent the number of loops in perturbation
theory.  At one-loop order, the value of $H$ is thus calculated by
\begin{equation}
H_{1} = H_{1} \ast f_{0} \ast \phi_{0} 
      = A_{1} - H_{0} \ast f_{1} \ast \phi_{0} 
              - H_{0} \ast f_{0} \ast \phi_{1} ,
\label{H1}
\end{equation}
the lowest order values $f_{0}$ and $\phi_{0}$ on quark states being
trivial delta functions.

Now, the factor of $1/\epsilon$ that gave the nonzero result for
$A_{1}$ is obtained from the very collinear regions that are to be
subtracted: The subtractions ensure that the hard-scattering
coefficient $H_1$ is dominated by the region where all the lines have
virtualities of order $Q^2$.  The collinear divergence in the
subtractions in Eq.\ (\ref{H1}) is multiplied by a factor proportional
to $\epsilon$, namely the lowest-order coefficient $H_0$, and one
obtains exactly the same factors of $1/\epsilon$ and $\epsilon$ as in
the corresponding contributions to $A_1$.  Thus the sum of the
subtraction terms cancels the finite nonzero term in $A_{1}$, so that
$H_{1}$ vanishes at $\epsilon = 0$.  We will show this in more detail
in Sec.~\ref{proof}.  When one takes parton distributions and
distribution amplitudes that are chirally even instead of chirally odd
then the corresponding $1/\epsilon$ factors are not multiplied by a
factor that vanishes at $\epsilon=0$, and in that case the subtraction
terms cancel the divergent terms so that $H_1$ is finite.  This
cancellation is guaranteed by the factorization theorem.

A clear symptom that the result of Ref.\ \cite{hl} cannot provide a
correct calculation of the coefficient function is that it comes from
a collinear region.\footnote{Hoodbhoy and Lu also obtain a finite term
  by multiplying the factor $\epsilon$ from the tree graphs with a
  $1/\epsilon$ ultraviolet pole from the wave function renormalization
  factors of the external quark lines.  However, if one works with
  renormalized amplitudes, as is always correct, the ultra-violet
  divergences are cancelled by renormalization counterterms. Moreover,
  no factors associated with propagator corrections on external lines
  need be considered when calculating a hard-scattering coefficient,
  since they identically cancel in the subtractions.}  Coefficient
functions, however, are collinear safe quantities.

Often one gets the impression from the literature that obtaining a
one-loop coefficient function is merely a matter of subtracting
$1/\epsilon$ poles from the unsubtracted graphs.  This is not the
case. Fundamentally, the correct subtractions have the form shown in
Eq.\ (\ref{H1}); this is necessary to avoid double counting
\cite{subtractions}.  Only if the Born graphs are independent of
$\epsilon$ is the subtraction formula (\ref{H1}) equivalent to the
subtraction of poles.  The difference between the correct and
incorrect procedures is, of course, particularly noticeable when the
Born graphs are proportional to $\epsilon$ so that there is no pole at
the one-loop level.

\section{General proof}
\label{proof}

In this section we close a loophole in the proof given in \cite{dgp}
that the coefficient function for the electroproduction of
transversely polarized vector mesons vanishes to all orders of
perturbation theory.  That proof does not treat the complications due
to the fact that chiral symmetry is broken in massless QCD whenever
the theory is regulated.

\begin{figure}
   \begin{center}
        \leavevmode
        \begin{tabular}{c@{~~~~~}c}
           \epsfxsize=0.4\hsize \epsfbox{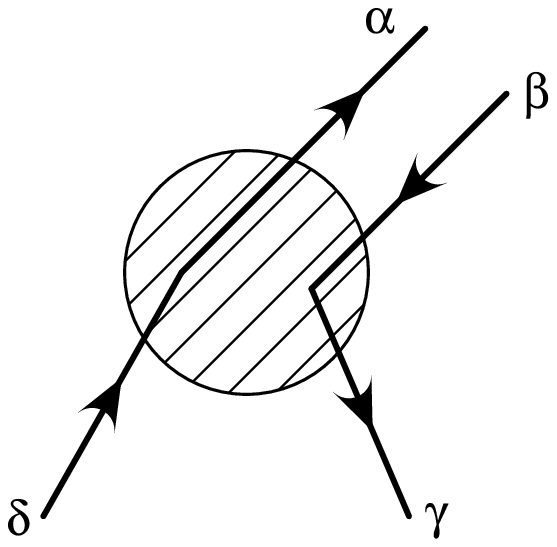} 
           & \epsfxsize=0.36\hsize\epsfbox{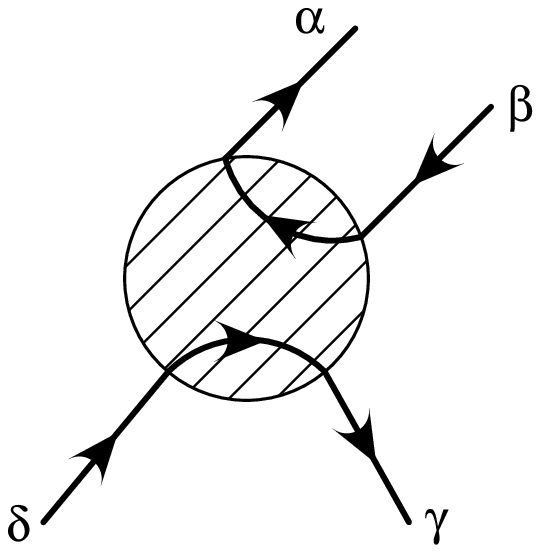}
        \\
                 (a)~~~~   &       (b)~~~
        \end{tabular}
   \end{center}
   \caption{Hard-scattering coefficient with spinor labels.  The
     external photon is not shown, and diagrams (a) and (b) show the
     two possibilities for connecting the external quark lines.}
   \label{fig.labels}
\end{figure}

With the spin-dependent factors for the external quarks removed, the
coefficient function has four Dirac indices:
$H_{\alpha}{}^\beta{}_{\gamma}{}^\delta$, as illustrated in Fig.\ 
\ref{fig.labels}.  The lower two quark lines are associated with the
proton, and have momenta in the $+z$ direction.  The upper two lines
are associated with the meson and have momenta in the $-z$ direction.
Since $H$ is a hard-scattering coefficient, all its external momenta
are massless and on-shell, and all internal masses are set to zero.

The theorem we will prove is that $H$ vanishes when contracted with
the Dirac matrices appropriate for a transversity distribution in the
proton and for a transversely polarized meson:
\begin{equation}
   H_{\alpha}{}^\beta{}_{\gamma}{}^\delta
   ~
   (\gamma^{+}\gamma^{i})_{\beta}{}^{\alpha}
   ~
   (\gamma^{-}\gamma^{j})_{\delta}{}^\gamma
   = 0.
\label{theorem}
\end{equation}
Here $i$ and $j$ are transverse indices associated with the
polarization states of the meson and proton.

There are two possibilities for the topological connection of the
external quark lines.  The first case is shown in Fig.\ 
\ref{fig.labels}(a), where the lines from the proton join to the lines
from the meson.  The second case is shown in Fig.\ 
\ref{fig.labels}(b), where the lines from the proton join to each
other and the lines from the meson join to each other.  As we will
discuss below, there is an odd number of Dirac matrices on each of the
internal lines of the coefficient function.  Since the trace of an odd
number of Dirac matrices vanishes, graphs of type (b) give zero when
contracted with the external Dirac matrices in Eq.\ (\ref{theorem}).
Hence only graphs of type (a) need to be considered.

\subsection{Previous proof}  
\label{previous.proof}

Ref.\ \cite{dgp} gave two proofs: One is that chirality is conserved
in massless QCD; this forces the initial and final quark states in
Fig.\ \ref{fig.labels} to have $J_{z}$ differing by two units, which
is prohibited by angular momentum conservation because the reaction is
collinear and the photon can at most provide one unit of $J_{z}$.  The
second proof, which translates this into algebra, is to observe that
the Feynman rules of massless QCD imply that
$H_{\alpha}{}^\beta{}_{\gamma}{}^\delta$ is the tensor product of two
fermion lines, each containing an odd number of Dirac matrices, as in
Fig.\ \ref{fig.LO} or \ref{fig.NLO}.  The only possibilities that have
the correct transformation under rotations about the $z$-axis give
Eq.\ (\ref{theorem}).

The first proof is incomplete as it does not discuss possible problems
caused by the anomalous breaking of chiral symmetry in massless QCD.

For the second proof we remark that the property of having an odd
number of Dirac matrices along each fermion line remains true in the
dimensionally regularized theory.  This is not changed by
manipulations such as decreasing the number of matrices by using $\{
\gamma^\mu, \gamma^\nu \} = 2 g^{\mu \nu}$ in an arbitrary number of
space-time dimensions.

Now, the calculation of Ref.\ \cite{hl} tells us that Eq.\ 
(\ref{theorem}) is false when applied to unsubtracted amplitudes. This
can be understood by observing that in order to use chirality
conservation and to perform the algebraic manipulations in the second
proof of Ref.\ \cite{dgp} one must be in 4 dimensions.  Without
subtractions, however, one cannot set $\epsilon=0$ in
$H_{\alpha}{}^\beta{}_{\gamma}{}^\delta$ because of the pole
$1/\epsilon$ from the momentum integration.  Hoodbhoy and Lu obtain a
finite result only after multiplying with the Dirac matrices for the
external lines, which provide a factor of $\epsilon$.  In this way
chirality conservation is broken because of non-trivial contributions
from the Dirac matrices in the unphysical
dimensions.\footnote{Remember that for $\mu\neq 0,1,2,3$ the matrices
  $\gamma^\mu$ do not anticommute with $\gamma_5 = i \gamma^0 \gamma^1
  \gamma^2 \gamma^3$.}

\subsection{Completed proof}
\label{completed.proof}

We now show that in the hard-scattering coefficient
$H_{\alpha}{}^\beta{}_{\gamma}{}^\delta$, once one has made the
appropriate subtractions and taken the limit $\epsilon=0$, chirality
is conserved to all orders in perturbation theory, so that both proofs
of Ref.\ \cite{dgp} apply and Eq.\ (\ref{theorem}) is correct.  We
start from the Feynman diagrams for the hard scattering, without
multiplying by the Dirac factors of the external quark lines, and
proceed in several steps.

\begin{enumerate}
\item Regulate the loop integrals by going to $4-\epsilon$ dimensions.
\item Perform the usual ultraviolet subtractions in the diagrams, as
  specified by the counterterms in the Lagrangian.  Soft or infra-red
  divergences cancel after summing over all graphs at a given order in
  $\alpha_{s}$ \cite{cfs}, so that after this step only collinear
  divergences are left.
\item Perform the subtractions for the collinear regions, as in Eq.\ 
  (\ref{H1}) and its generalizations to higher order.  Note that the
  subtraction scheme is defined by the {\em ultra-violet}
  renormalization that is used to define the parton densities.  As in
  step 2, one has a choice of subtraction scheme and may for instance
  use the $\overline{\rm MS}$ prescription.
\item Remove the regulator, i.e., take the limit $\epsilon=0$.  The
  crucial fact is that one obtains a finite result for
  $H_{\alpha}{}^\beta{}_{\gamma}{}^\delta$ before contracting with any
  Dirac matrices.  This follows from the general construction
  of the factorization proof \cite{cfs}: Momentum regions for Feynman
  graphs for the overall process are characterized by a division into
  subgraphs associated with the three building blocks $H$, $f$ and
  $\phi$ of the factorization formula (\ref{factorize}).  To show that
  the hard-scattering coefficient $H$ is finite and dominated by large
  virtualities does not require use of the Dirac algebra.  Hence we
  can set $\epsilon=0$ before projecting out any particular Dirac
  structure. 
\end{enumerate}
None of these steps impinges on the fact that for each quark line we
have a string of an odd number of Dirac matrices.  This is ensured by
the vector coupling structure of the Lagrangian and of the
counterterms.  Now, a product of an odd number of basic Dirac matrices
in 4 dimensions conserves the chirality of the external lines, from
which follows the same property for $H$.  This completes the proof.

Let us illustrate these steps by the following integral that contains
a typical structure in the one-loop contribution to
$H_{\alpha}{}^\beta{}_{\gamma}{}^\delta$.  Without subtractions it is
\begin{equation}
   I_{\alpha}{}^{\beta}{}_\gamma{}^{\delta}
   = (\gamma^{\lambda})_{\gamma}{}^{\beta}
     (\gamma_{\lambda})_{\alpha}{}^{\delta} \;
     \mu^{\epsilon} 
     \int \frac{d^{2-\epsilon}k_{T}}{(2\pi)^{2-\epsilon}}
     ~ \frac{f(k_{T})}{k_{T}^{2}} .
\label{I}
\end{equation}
This contains an integral over transverse momentum which becomes
divergent as $k_{T} \to 0$ when $\epsilon = 0$.  The factor $f(k_{T})$
represents the rest of the integrand and provides a cut-off at large
$k_{T}$, so that there are no ultra-violet divergences in this case.
The contribution appropriate for the coefficient of a transversely
polarized meson is obtained by contracting with
$(\gamma^{+}\gamma^{i})_{\beta}{}^{\alpha}$; then the Dirac algebra
gives a factor proportional to $\epsilon$.

After subtraction of the collinear divergence we obtain
\begin{eqnarray}
    \overline{I}_{\alpha}{}^\beta{}_{\gamma}{}^\delta 
   &=& (\gamma^{\lambda})_{\gamma}{}^{\beta}
       (\gamma_{\lambda})_{\alpha}{}^{\delta}
       \left\{
            \mu^{\epsilon} 
            \int \frac{d^{2-\epsilon}k_{T}}{(2\pi)^{2-\epsilon}}
            \left [ \frac{f(k_{T})}{k_{T}^{2}} 
             - \frac{f(0)}{k_{T}^{2}}
            \right ]
       \right.
\nonumber\\
    && ~~~~~~~~~~~~~~~~~
         \left.
             + \frac{f(0)}{4\pi}
               \left( \frac{2}{\epsilon} + \ln 4\pi - \gamma
               \right) 
       \right\} .
\label{I.bar}
\end{eqnarray}
Here the second term inside the square brackets represents an
integrand for the parton density or the meson distribution amplitude.
It cancels the singularity in the integrand but introduces an
ultra-violet divergence that we renormalize by the last term, which is
an $\overline{\rm MS}$ counterterm. The factor that multiplies the
Dirac matrices, i.e., the factor in braces, is finite as $\epsilon \to
0$.

A convenient way \cite{jcc1} of representing the counterterm is as an
integral over $k_T$, so that at $\epsilon=0$ we have a finite integral
for $\overline{I}$:
\begin{eqnarray}
    \overline{I}_{\alpha}{}^\beta{}_{\gamma}{}^\delta 
   &=& (\gamma^{\lambda})_{\gamma}{}^{\beta}
       (\gamma_{\lambda})_{\alpha}{}^{\delta}
\nonumber\\
   && ~~~~
            \int_0^\infty \frac{dk_{T}^2}{4\pi}
            \left [ \frac{f(k_{T})}{k_{T}^{2}} 
             - \frac{f(0)}{k_{T}^{2}} \theta (\mu - k_{T})
            \right ] .
\label{I.bar2}
\end{eqnarray}
Clearly, we now have 4-dimensional Dirac matrices multiplying a finite
integral.  This is the final result for the contribution to $H$, after
step 4 of the construction, and chirality conservation clearly holds.

A few remarks are in order about multiplying with the Dirac matrices
for the external lines.  Clearly this multiplication must \emph{not}
be done before regularization, otherwise one risks multiplying an
infinity from the loop integrals by a zero from the Dirac algebra, and
thereby losing control over the calculation.  It \emph{can} be done
after step 1 as all expressions are well defined, and if one correctly
performs all subtractions, one will obtain the correct result.
However, as explained at the end of Sec.~\ref{previous.proof}, it is
difficult to keep track of chirality in $4-\epsilon$ dimensions, where
this symmetry is broken.  Postponing the multiplication by external
factors until after step 4 has been performed enables us to show that
no chirality breaking effects from the extra dimensions survive in the
hard-scattering coefficient.

Dimensional regularization is not the only possibility to regulate the
collinear divergences in the hard scattering.  For an alternative
proof one could let the external quark lines in $H$ be slightly
off-shell. (One may want to do this without giving them transverse
momentum so as to keep rotation invariance in the $x$-$y$ plane, which
is the other essential ingredient in the proof next to chirality
conservation.)  Then one can take the limit $\epsilon=0$ already after
step~2.  Removing the regulator in step 4 then means that we put the
external lines back on shell.

Clearly our arguments leading to chirality conservation in the
hard-scattering coefficient are not specific to exclusive vector meson
production, and generalize to other processes.

\section{The axial anomaly}
\label{anomaly}

Let us make some remarks about how one can understand our result in
the light of the axial anomaly in QCD.  We have shown that chirality
is conserved for the coefficient function
$H_{\alpha}{}^\beta{}_{\gamma}{}^\delta$, i.e., that it is invariant
under a chiral transformation of the external quark legs:
\begin{equation}
H_{\alpha}{}^{\beta}{}_{\gamma}{}^{\delta} =
U_{\alpha}{}^{\alpha'}\,
U_{\gamma}{}^{\gamma'}\,
H_{\alpha'}{}^{\beta'}{}_{\gamma'}{}^{\delta'}\,
U_{\beta'}{}^{\beta}\,
U_{\delta'}{}^{\delta}
\label{chiral}
\end{equation}
where $U = \exp(i \omega \gamma_5)$ with a real parameter $\omega$.
The invariance follows simply from the fact that $H$ is a tensor
product of strings of an odd number of basic 4-dimensional Dirac
matrices (i.e., of $\gamma^0$, \dots, $\gamma^3$).

To obtain this result we used arguments about the counting of Dirac
matrices that remain valid in the dimensionally regulated theory.
However, we only invoke chirality once we are back in 4 dimensions.
We have not made use of the Noether theorem and the axial current or
axial charge.  To prove the Ward identity associated with chiral
symmetry requires more than we needed in our proof, and, as is well
known, the proof that the axial current is conserved fails.

Even though the axial anomaly in QCD can be calculated in one-loop
perturbation theory, its existence does not of itself imply broken
chiral symmetry.  The anomaly is merely an enabling result that
permits anomalous chiral symmetry breaking.  It tells us that the
divergence of the axial current, $\partial \cdot j_5$, is another
operator, proportional to $F \cdot \tilde F$, but it does not tell us
that matrix elements of this operator are non-zero.  Note that by its
definition $F \cdot \tilde F$ is a pure divergence, although of a
non-gauge invariant operator.  The anomaly is a necessary but not
sufficient condition for the breaking of chiral symmetry.  As has been
observed in other work \cite{chiral.breaking}, the actual breaking of
chiral symmetry in QCD is purely non-perturbative.  There is no
anomalous breaking of chiral symmetry in any finite order of
perturbation theory, provided that a genuinely infrared safe quantity
is calculated: hard-scattering coefficients respect chiral invariance
in the sense of Eq.\ (\ref{chiral}).  Notice that in the context of
the spin crisis the axial anomaly appears in the hadronic matrix
elements that define the parton densities, i.e.,\ in non-perturbative
quantities.

\section{Conclusions}
\label{concl}

We have shown explicitly that chiral symmetry for massless external
fermion lines is preserved order by order in perturbative QCD,
provided that an infrared safe quantity is calculated.  This implies
that, in the electroproduction of mesons, the hard-scattering
coefficient vanishes when one takes the component that couples to a
transversity distribution.  Thus, the proposal of Collins, Frankfurt
and Strikman \cite{cfs} to use this process for measuring transversity
distributions does not work.  The claim of Hoodbhoy and Lu \cite{hl}
to the opposite effect fails because the authors omitted to make the
collinear subtractions necessary in a correct calculation of a
hard-scattering coefficient.  Our proof completes that given by Diehl,
Gousset and Pire \cite{dgp}.

The methods we employ apply to other cases, and show that the
application of chiral symmetry to hard-scattering coefficients is
generally valid.

Our result is in one sense negative, since it shows that
electroproduction of mesons cannot be used to measure transversity
distributions.  However it also provides a helicity selection rule: If
any exclusive electroproduction of transversely polarized vector
mesons is observed, it must be due to a power correction.  In this
respect the process is very similar to exclusive processes where
hadron helicity conservation holds, such as electron-hadron or
hadron-hadron scattering at large momentum transfer, or to the
structure function $g_2$, which provides a probe of higher-twist
physics uncontaminated by leading twist.

\section*{Acknowledgments}

This work was supported in part by the U.S.\ Department of Energy
under grant number DE-FG02-90ER-40577, and in part by the TMR program
of the European Union under contract number FMRX-CT98-0194.  M.D.
would like to thank M. Beneke, T. Gousset, B. Pire and A.V. Radyushkin
for discussions and E. Pilon for correspondence.  J.C. would like to
thank M.~Eides for discussions.

\end{document}